\documentclass[eqsecnum,aps,twocolumn,epsf,showpacs]{revtex4} 
\usepackage{graphicx}
 
\begin{document}
   
 


\title{Black soliton in a quasi-one-dimensional trapped 
fermion-fermion mixture}

 
\author{Sadhan K. Adhikari\footnote{Electronic
address: adhikari@ift.unesp.br; \\
URL: http://www.ift.unesp.br/users/adhikari/}
}
\affiliation
{Instituto de F\'{\i}sica Te\'orica, UNESP $-$ S\~ao Paulo
State University,  01.405-900 S\~ao Paulo, S\~ao Paulo, Brazil}
 
\date{\today}
 
 
\begin{abstract}
Employing a time-dependent mean-field-hydrodynamic model we
study the generation of black solitons in a degenerate fermion-fermion
mixture in a cigar-shaped geometry using variational and numerical
solutions. The black soliton is found to be the first stationary
vibrational excitation of the system and is considered to be a nonlinear
continuation of the vibrational excitation of the harmonic oscillator
state. We illustrate the stationary nature of the black soliton, by
studying different perturbations on it after its formation.


\end{abstract}



\maketitle

\section{Introduction}

The one-dimensional nonlinear Schr\"odinger (NLS) equation in the
repulsive or self-defocusing case
is usually written as \cite{1} \begin{equation}\label{nls} i u_t+u_{xx}-
|u|^2u=0, \end{equation} where the time ($t$) and space ($x$) dependences
of the wave function $u(x,t)$ are suppressed. This equation sustains the
following dark and grey solitons \cite{5}: \begin{eqnarray}\label{DS}
u(x,t)=r(x-ct) \exp[-i\{\phi(x-ct)-\mu t \}], \end{eqnarray} with
\begin{eqnarray} r^2(x-ct)& = & \eta -2\xi^2 \mbox{sech}^2[\xi(x-ct)], \\
\phi(x-ct)&=&\tan^{-1}[-2 \xi/c \hskip 0.05cm \tanh\-\{\xi (x-ct)\}], \\
\xi
&=& \sqrt{(2\eta - c^2)}/2, \end{eqnarray} where $c$ is the velocity,
$\mu$ is the chemical potential, and $\eta$ is related to
intensity. Soliton
(\ref{DS}) having a ``notch" over a background density is grey in general.
It is dark if density $|u|^2=0$ at the minimum. At zero velocity the
soliton becomes a stationary dark  soliton or a black soliton: $|u(x,t)|=
\sqrt {\eta} \tanh
[x\sqrt{\eta/2}]$.

The similarity of the NLS equation (\ref{nls}) to the
mean-field Gross-Pitaevskii
(GP)
equation describing a 
trapped zero-temperature Bose-Einstein condensate (BEC) \cite{rmp}  
imply the possibility of a stationary
dark soliton, or a black soliton, in a trapped BEC or in
a degenerate trapped boson-fermion mixture (DBFM)
\cite{bur}.  Dark solitons represent a 
local minimum in a trapped BEC and have been the topic of both
experimental  
\cite{expds} and theoretical 
\cite{bur,ds1,ds2} investigations.
Both DBFM \cite{expbf} and degenerate fermion-fer\-mion
mixture
(DFFM) \cite{expff}
have been experimentally observed and studied theoretically
\cite{yyy,capu}. 
It has been suggested that the black soliton of a harmonically
trapped BEC
could be a stationary eigenstate \cite{bur} of the GP
equation \cite{rmp} as in
the case of the trap-less NLS equation.  
We demonstrated  that 
the black soliton in a trapped zero-temperature BEC \cite{dsa}
or a DBFM
\cite{fds}
is the first stationary vibrational excitation of
the nonlinear dynamical equation for the respective systems.
No such vibrational excitation 
of the NLS equation (\ref{nls}) exists in the absence of a harmonic
 trap.

In this paper we study the numerical simulation of a black or a stationary
dark soliton in a trapped DFFM.
We use a coupled time-dependent mean-field-hydrodynamic model for a DFFM
and consider the formation of black  solitons in a
quasi-one-dimensional cigar-shaped geometry using numerical and
variational solutions.  The present model is inspired by the success of a
similar model used recently in the investigation of collapse \cite{ska},
bright \cite{fbs2} and dark \cite{fds} solitons in a DBFM and 
of mixing-demixing in a DFFM \cite{mix}. The black
soliton of a DFFM is demonstrated to be a stationary vibrational
excitation of the mean-field-hydrodynamic  model.

We suggest a scheme for numerical simulation
of a stationary dark soliton in trapped DFFM
by time evolution of the linear harmonic oscillator equation 
starting with the analytic vibrational excitation, 
while the nonlinearities corresponding to the mean-field-hydrodynamic
equations
are slowly introduced.  
The simulation proceeds through successive eigenstates of
the dynamical model.   To illustrate the
stability of our scheme we study  the oscillation of the
stationary dark soliton upon application of different perturbations
\cite{dsa}.

Collective excitations in the form of solitons and vortices in trapped
fermions have also been investigated recently by Damski et al. \cite{x}
and Karpiuk et al. \cite{y}. However, these authors  considered isolated
ultra-cold
fermions and not a realistic DFFM as in the
experiments and as discussed in this paper.  Also they did not demonstrate
the existence of stable dark solitons with a zero at the center of the
notch as in the present study. They identified grey soliton-like structure
with a shallow dip in the isolated fermionic density distribution quite
distinct from the stable fermionic dark solitons in a DFFM as
noted in this investigation.

Though the present dark solitons are stable in the mean-field
formulation, they could be unstable physically due to quantum fluctuations
\cite{z}. The effect of quantum fluctuations is lost in the mean-free
model and can only be studied in a field-theoretic approach. Moreover
being an excited state they are thermodynamically unstable.  There have
been suggestions about how to excite a dark soliton by phase imprinting
method \cite{x,y}.  The dark soliton is the lowest vibrational excitation
of the mean-field model \cite{dsa} and there have also been investigations
about how to
attain such excited states \cite{yu}.  Nevertheless, despite different
suggestions about how to excite a dark soliton \cite{x,y,yu}, experiments
to date have not yet generated a stationary dark soliton.
Experimentally, so far the dark solitons have been unstable \cite{bur}.
However, considering that they are stationary excitations
of the mean-field-hydrodynamic  equation their creation {\it must} be
possible 
at
least as a
non-stationary dark soliton which may turn grey and oscillate before
decaying due to quantum fluctuations and thermodynamic effects.

The plan of the paper is as follows.
In Sec. II we present the
mean-field-hydrodynamic model 
which we use in the present study. We also perform a reduction
of the model equations to a quasi-one-dimensional form 
in a cigar-shaped geometry. 
In Sec. III we present a variational analysis
of the mean-field equations. In Sec. IV we present the numerical
results for stationary dark solitons
and
compare
with the variational results. 
The stationary nature of the dark
solitons is illustrated  through a  numerical study of their 
dynamics  under different perturbing forces. 
Finally, we present a summary in
Sec. V.

\section{Mean-field-hydrodynamic Model}

To  develop a set of practical time-dependent
mean-field-hydrodynamic
equations for a DFFM, we consider   the
following Lagrangian density \cite{fbs2,fds}
\begin{eqnarray}\label{yy}
&{\cal L}& = \frac{i}{2}\hbar \sum_{j=1}^2\left(
\Psi_j \frac{\partial \Psi_j \- ^*}{\partial t} - \Psi_j \- ^*
\frac{\partial \Psi_j
}{\partial
t}
\right)+ g_{12}  n_1n_2 \nonumber \\
&+& \sum_{j=1}^2 \left(\frac{\hbar^2 |\nabla
  \Psi_j |^2 }{6m_j}+
V_j n_j+\frac{3}{5}  A_j  n_j ^{5/3}\right),
\end{eqnarray}
where  $m_j$ is the
mass 
of component $j (=1,2)$, $\Psi_j$ a complex probability amplitude,
$n_j=|\Psi_j|^2$ the real probability density,
$N_j \equiv  \int d{\bf r} n_j({\bf r}) $  the
number of respective atoms,
$A_j =\hbar^2(6\pi^2)^{2/3}$ $/(2m_j)$, and $V_j$ is the confining
trap.
Here the  interspecies
coupling is
$g_{12}=2\pi \hbar^2 a_{12}
/m_R$
with the 
reduced mass $m_R=m_1m_2/(m_1+m_2),$ and  $ a_{12}$
is the interspecies
scattering length.
The interaction between intra-species fermions in
spin-polarized state is highly suppressed due
to the Pauli blocking  term $3A_jn_j^{5/3}/5$
and has been neglected in  Eq. (\ref{yy})  and will be
neglected throughout.
The kinetic energy terms in this equation  $\hbar^2|\nabla
\Psi_j|^2/(6m_j)$
are derived from a hydrodynamic equation for the
 fermions
\cite{capu} and  contribute little to this problem compared to the
dominating Pauli blocking term   in  Eq. (\ref{yy}).
However, the  inclusion of the kinetic energy terms in  Eq. (\ref{yy})
leads
to a smooth  solution for the probability density everywhere
\cite{fbs2}. To keep the algebra simple and
without losing generality,  in our calculation  we shall
take
equal
fermion masses: 
$m_1=m_2\equiv m/3.$ This simulates well the fermion mixtures in two
hyperfine  states of $^6$Li  and $^{40}$K 
atoms observed
experimentally \cite{expff}.
The Lagrangian density of each fermion component in
Eq. (\ref{yy}) is
identical to that used in Refs.  \cite{fds,fbs2}.

The mean-field dynamical equations for the system are just the
usual  Euler-Lagrange  (EL) equations \cite{gold}
\begin{equation}
\frac{d}{dt}\frac{\partial {\cal L}}{\partial   \frac{\partial
\Psi_j^*}{\partial t}}+
\sum _{k=1}^3 \frac{d}{dx_k}\frac{\partial {\cal L}}{\partial
\frac{\partial \Psi_j^*}{\partial x_k}}= \frac{\partial {\cal
L}}{\partial
 \Psi_j^*},
\end{equation}
where $x_k, k=1,2,3,$ are the three space components. 
With Lagrangian density
(\ref{yy}) the following  EL   equations of
motion are
derived in a straight-forward fashion \cite{ska,fbs2}:
\begin{eqnarray}\label{e}  \biggr[ - i\hbar\frac{\partial
}{\partial t}
-\frac{\hbar^2\nabla_{\bf r}^2}{6m_{{j}}}
+ V_{{j}}({\bf r})+A_j n_j^{2/3}
&+& g_{{12}} n_k
 \biggr]\Psi_j=0, \nonumber \\
 j \ne k=1,2.  \end{eqnarray} 
This is 
essentially a time-dependent version
of a similar time-independent model used recently for 
fermions \cite{md2}.
For a stationary state Eqs. (\ref{e}) yield the same result as the
formulation of Ref. \cite{capu,md2}. For a system with large number 
of fermions
both reduce to \cite{ska,md2} the well-known Thomas-Fermi 
approximation
\cite{rmp,str}:
$n_j=[(\mu_j -V_j)/A_j]^{3/2}$, with $\mu_j$ the chemical potential.

We  reduce  Eq. (\ref{e})  to a  minimal
 quasi-one-dimensional form
in a cigar-shaped geometry  where the confining trap of anisotropy
$\nu$
has the form
 $V_j({\bf r})= \frac{1}{2}3 m_j\omega^2(
\rho^2+\nu^2 z^2),$ with $\rho$ the radial vector and $z$ the axial
vector.
For  a cigar-shaped geometry  $\nu << 1$,
we consider solutions of    Eq.   (\ref{e})  of the type
$\Psi_j({\bf r},t)= \sqrt N_j \phi_j(x,\tau)\psi_j^{(0)}(
\rho)/\sqrt{l},
$
where $x=z/l$, $\tau=t\nu \omega/2$, $l=\sqrt{\hbar/(\nu \omega m)}$
and $\psi_j^{(0)}( \rho)$ is the circularly symmetric ground-state linear
harmonic-oscillator wave function. In the $\nu << 1$ limit the
quasi-one-dimensional 
functions $\phi_j(x,\tau)$ satisfy the following dimensionless
equations \cite{mix}:
\begin{eqnarray}\label{n}
 \biggr[ &-&  i\frac{\partial
}{\partial \tau}-\frac{\partial ^2}{\partial x^2}
+ x^2     +
{\cal N}_{jk}
  \left|{{\phi}_k} \right|^2\nonumber \\
&+&
{\cal N}_{jj}
  \left|{{\phi}_j} 
\right|^{4/3}
 \biggr]{\phi}_{{j}}(x,\tau)=0,\quad
 j\ne k =1,2,
\end{eqnarray}
where
${\cal N}_{jk}=12 a_{12}(N_k/\nu)/l,$ and
${\cal N}_{jj}=9(6\pi N_j/\nu)^{2/3}/5. $ 
In Eq. 
 (\ref{n}),
the normalization condition  is given by
$\int_{-\infty}^\infty \-|\phi_j(x,\tau)|^2 dx =1 $.

In case when $N_1=N_2=N$, the two equations  (\ref{n}) become identical
and reduce to 
\begin{eqnarray}\label{o}
 \biggr[ -  i\frac{\partial
}{\partial \tau}-\frac{\partial ^2}{\partial x^2}
+ x^2     +
\gamma 
  \left|{{\phi}} \right|^2
+
\beta 
  \left|{{\phi}} 
\right|^{4/3}
 \biggr]{\phi}(x,\tau)=0, \nonumber \\
\end{eqnarray}
which 
permits the solution $\phi=\phi_1=\phi_2$, with 
 $\beta = {\cal N}_{jj}$ and $\gamma= 
 {\cal N}_{jk}$.

\section{Variational Analysis}

\subsection{Symmetric Case ($N_1=N_2$)}

Next we present a 
variational analysis of Eq. (\ref{o}) based on the stationary
trial wave function \cite{and}
\begin{equation}\label{v}
\phi_v(x,\tau)=A(\tau)  x\exp\biggr[
-\frac{x^2}{2R^2(\tau)}+\frac{i}{2}
b(\tau)x^2+ic(\tau) \biggr], 
\end{equation}
where $A$ is the amplitude,  $R$ is the width,   $b$ the chirp, and $c$
the phase. 
The Lagrangian density for Eq. (\ref{o})
is the one-term version of Eq. 
(\ref{yy}), e.g., 
\begin{eqnarray}
{\cal L} &=& \frac{i\hbar}{2} \biggr[\phi_v
\frac{\partial \phi_v^*
}{\partial t } - \phi_v ^* \frac{\partial \phi_v
}{\partial t }
\biggr]+ \biggr|\frac{\partial  \phi_v}{\partial x} \biggr|^2
+\frac{1}{2}
\gamma n ^2 
+ \frac{3}{5}\beta n^{
5/3}\nonumber \\
&+& x^2 n
\end{eqnarray}
 which is evaluated with this 
trial function and the effective
Lagrangian  $L
=\int_{-\infty}^\infty {\cal L}(\varphi_v)dx$ becomes 
\begin{eqnarray}
L &=& \frac{A^2R^3\sqrt \pi}{2}
\biggr(
\alpha
\beta  (AR)  ^{4/3}+\dot c
+ \frac{3R^2\dot b}{4}
+\frac{3\gamma }{16\sqrt 2}  A ^2R^2          
\nonumber \\ & +&\frac{3}{2R^2}
+\frac{3}{2}b^2R^2 +\frac{3}{2}R^2
\biggr),
\end{eqnarray}
where $\alpha = ({54}/{125}\sqrt \pi) ({3}/{5})^{1/6} 
\Gamma(13/6)\approx 0.242269  $ and 
the overhead dot denotes time derivative.
The  variational Lagrange
equations \cite{gold}
\begin{equation}\label{la}
\frac{d}{dt}\frac{\partial L}{\partial \dot q}= \frac{\partial
L}{\partial q},
\end{equation}
where $q$ stands 
for $c$, $A$,   $R$ and $b$
can then be written as 
\begin{eqnarray}\label{con}
\frac{\sqrt \pi}{2}A^2 R^3 
=\mbox{constant}=1.
\end{eqnarray}
\begin{eqnarray}
3\dot b&=& -\frac{6}{R^4}-6b^2- \frac{3\gamma A^2}{2\sqrt 2}-6
-\frac{20}{3}\alpha \beta \frac{A^{4/3}}{R^{2/3}}-\frac{4\dot
c}{R^2},\label{x}\\
5\dot b&=& -\frac{2}{R^4}-10b^2-\frac{5\gamma A^2}{4\sqrt 2}-10
-\frac{52}{9}\alpha \beta \frac{A
^{4/3}}{R^{2/3}}-\frac{4\dot c}{R^2},\nonumber \\ \label{y}\\
\dot R &=&  2 R b.\label{z}
\end{eqnarray}
The constant in Eq. (\ref{con}) is fixed by the normalization condition. 
Eliminating $\dot c$ from 
Eqs. (\ref{x}) and (\ref{y}) we obtain 
\begin{eqnarray} \label{eli}
2\dot b = \frac{4}{R^4}-4b^2+\gamma\frac{A ^2}{4\sqrt 2}
-4
+\frac{8}{9} \alpha \beta \frac{A^{4/3}}{R^{2/3}}. 
 \end{eqnarray}      The use of 
Eqs. (\ref{con}), (\ref{z}) and (\ref{eli}) 
 leads to the following  
differential equation for the width $R$:
\begin{eqnarray} \label{min}
\frac{d^2 R}{d\tau^2}& =& 
\biggr(
\frac{4}{R^3}+ \frac{\gamma}{2\sqrt{2\pi}R^2} -4R
+\frac{8}{R^{5/3}}\frac{ 2^{2/3}\alpha \beta 
}{9\pi^{1/3}
} \biggr),\\
&=&-\frac{d}{dR}\biggr[  \frac{2}{R^2}
+\frac{\gamma}{R}
\frac{1}{2\sqrt{2\pi}}      +2R^2
+\frac{4\alpha \beta }{3} \left(\frac {4}{\pi R^2}
\right)^{1/3}
\biggr]. \nonumber \\
\end{eqnarray}
The quantity in the square bracket is the effective potential of the
equation of motion. Small oscillation around a stable configuration is
possible when there is a minimum in this potential. 
 The variational result for width $R$ follows by setting the
right-hand-side of Eq. (\ref{min}) to zero corresponding to a minimum in
this effective potential,  from which the variational 
profile for the soliton can be obtained \cite{and} using Eq. (\ref{v}). 

\subsection{Asymmetric Case ($N_1\ne N_2$)}

The above variational analysis can be extended to the asymmetric
case. However, the algebra becomes quite involved if we take a general
variational trial wave function with  chirp and phase parameters. As we
are interested mostly in the density profiles, we consider the following 
normalized stationary  trial wave function for fermion component $j$ of
Eq. (\ref{n})
\begin{equation}\label{vx}
\varphi_{vj}= \sqrt{\frac{2}{R_j^3(\tau)\sqrt \pi}}x \exp\biggr[
-\frac{x^2}{2R_j^2(\tau)}
\biggr], \quad j=1,2. 
\end{equation}
Using essentially the Lagrangian density (\ref{yy}) in this case we 
obtain the following effective Lagrangian 
\begin{eqnarray}
L=\frac{3N_{jk}}{\sqrt \pi}\frac{R_1^2R_2^2}{(R_1^2+R_2^2)^{5/2}}
+\sum_{j=1}^2\biggr(\frac{\alpha 2^{2/3} }{ 
\pi^{1/3}}\frac{N_{jj}}{R_j^{2/3}}+\frac{3(R_j^4+1)}{2R_j^2}
\biggr).  \nonumber \\
\end{eqnarray}
The variational Lagrange equations (\ref{la}) for $R_1$ and $R_2$ now
become
\begin{eqnarray}\label{cp}
\frac{4}{R_j^3}-4R_j -\frac{4N_{jk}}{\sqrt
\pi}\frac{2R_jR_k^4-3R_j^3R_k^2}{(R_1^2+R_2^
2)^{7/2}}
+\frac{2^{2/3}}{9 \pi ^{1/3}}\frac{8\alpha N_{jj}}{R_j^{5/3}}=0.
\nonumber \\
\end{eqnarray}
Equations (\ref{cp}) can be solved for the variational widths $R_j$ and
consequently the variational profile of the wave functions obtained 
from Eq. (\ref{vx}). When $N_1=N_2$, in Eq. (\ref{cp})  $N_{jj}= \beta$,  
$N_{jk}=N_{kj}=\gamma$ and $R_1=R_2=R$; and in this case it is verified
that 
Eq. (\ref{cp}) yields the same 
variational widths as from  the result obtained in the symmetric case in
Sec. IIIA given by 
Eq. (\ref{min}); e. g.,  
\begin{equation}\label{sim}
\frac{4}{R^3}-4R+\frac{\gamma}{R^2} 
{\frac{1}{2\sqrt{2\pi}}}+
\frac{2^{2/3}}{R^{5/3}}\frac{8 \beta\alpha }{9 
\pi^{1/3}}=0.
\end{equation}

\subsection{Black Soliton in a BEC}

By a variational calculation we 
establish the stability of a black soliton
in a trapped 
BEC satisfying the one-dimensional GP equation:
\begin{equation}\label{p}
 \biggr[ -  i\frac{\partial
}{\partial \tau}-\frac{\partial ^2}{\partial x^2}
+ x^2     +n
  \left|{{\phi}} \right|^2
 \biggr]{\phi}(x,\tau)=0.
\end{equation}
This equation is identical with Eq. (\ref{o}) with 
$\beta =0$ and $\gamma=n$. Hence the variational analysis of Sec. IIIA
applies to this
case and the differential equation for width $R$ is given by 
Eq.  (\ref{min}) with
$\beta =0$: 
\begin{equation}\label{q}
\frac{d^2 R}{d\tau^2} = 
\biggr(
\frac{4}{R^3}+ \frac{n}{2\sqrt{2\pi}R^2} -4R
 \biggr).
\end{equation}
The variational width in this case is given by setting 
the right-hand-side of this equation to zero.

\begin{figure}
 
\begin{center}
\includegraphics[width=.48\linewidth]{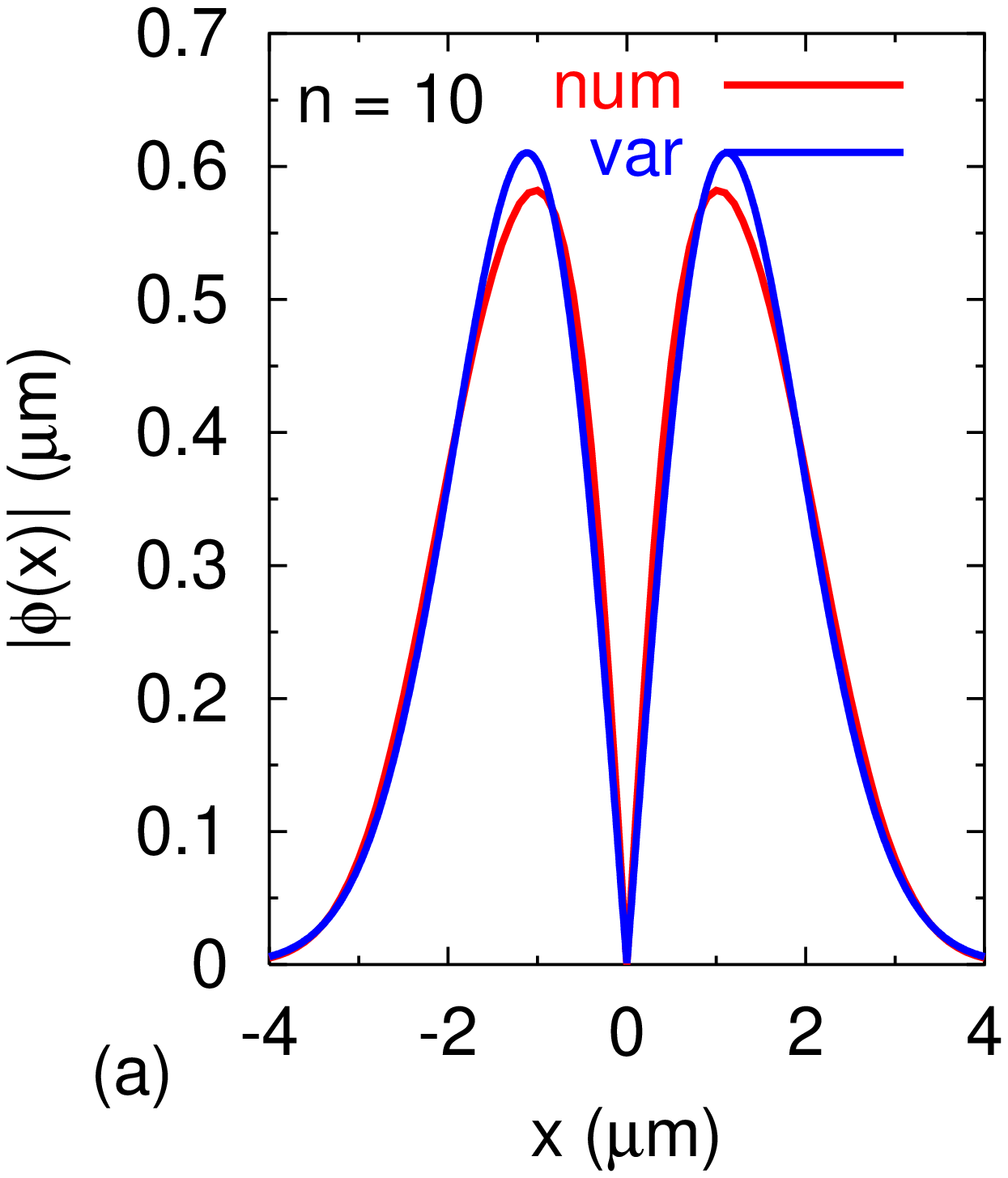}
\includegraphics[width=.48\linewidth]{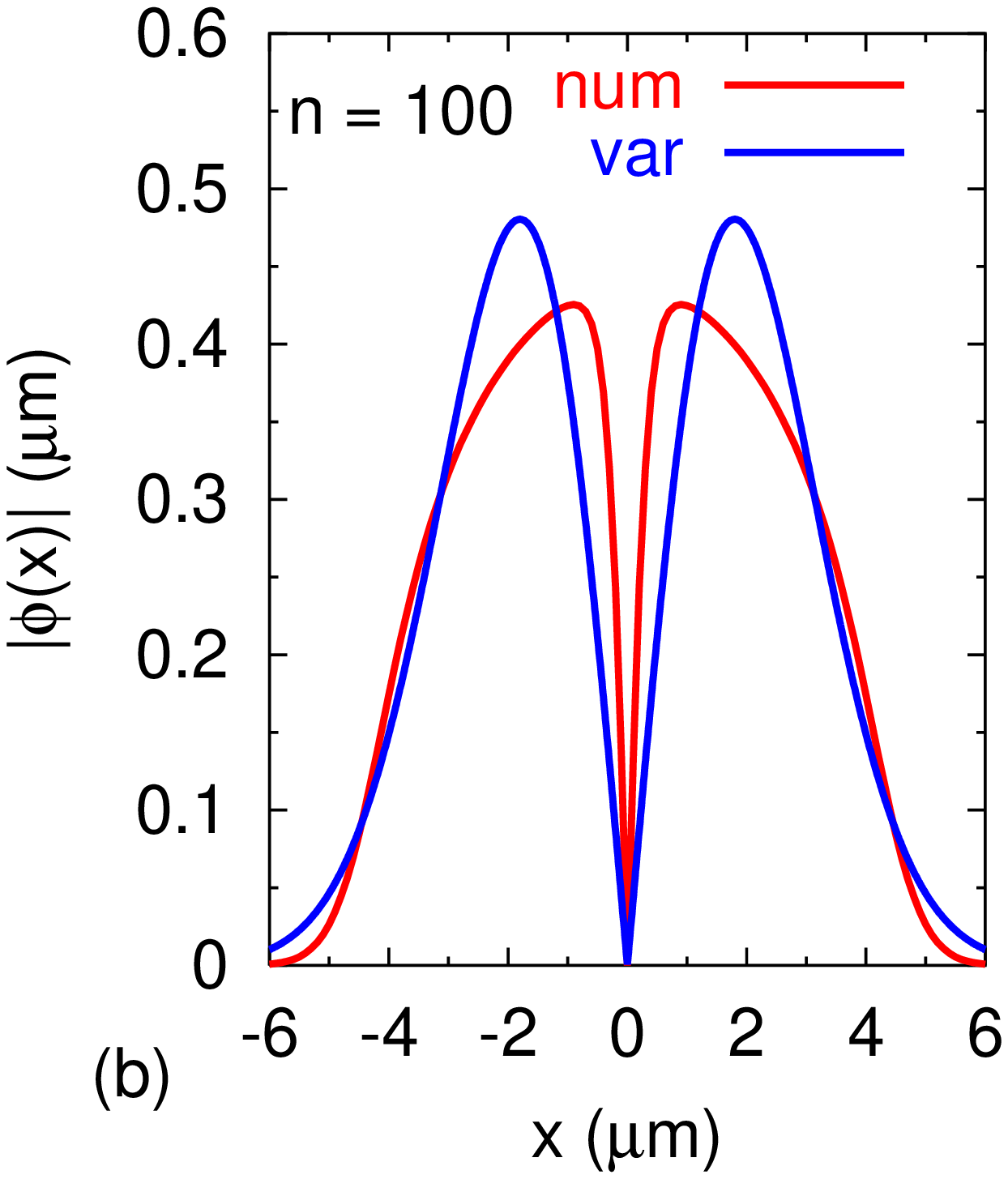}
\end{center}

\caption{(Color online)  The 
solitons  $|\phi(x)|$ of BEC from the numerical solution  of Eq. (\ref{p})
(denoted num)  vs. $x$ 
for  (a) $n=10$ and (b) $n=100$. 
The corresponding variational solutions 
obtained  from 
Eq. (\ref{q}) (denoted var) 
  are also shown. 
} \end{figure}

 \section{Numerical Results}

We solve Eqs.  
 (\ref{n})  for bright  solitons
numerically using a time-iteration
method based on the Crank-Nicholson discretization scheme
elaborated in Ref. \cite{sk1}. We discretize the coupled partial
differential equations  (\ref{n})    
using time step $0.0004$ and space step $0.02$ in the domain 
$-16<x<16$. The second derivative in $x$ is discretized by a three-point
finite-difference rule and the first derivative in time  by a two-point
finite-difference rule.  We start  a time evolution of Eqs.  (\ref{n}) 
setting the nonlinear terms to
zero, and  
starting with the eigenfunction of the linear harmonic
oscillator problem: $\phi(x,\tau) 
=\sqrt{(2/\sqrt \pi)}
x\exp(-x^2/2)$ $\exp(-3i\tau).$  
During the time evolution the nonlinear
terms are  switched on  slowly and  
the time evolution  continued to obtain the final converged
solutions.  
In addition to solving the coupled equations 
(\ref{n}), we also solved the single equation (\ref{o}) in the symmetric
case: $N_1=N_2$ and Eq. (\ref{p}) for bosons.

\begin{figure}
 
\begin{center}
\includegraphics[width=.8\linewidth]{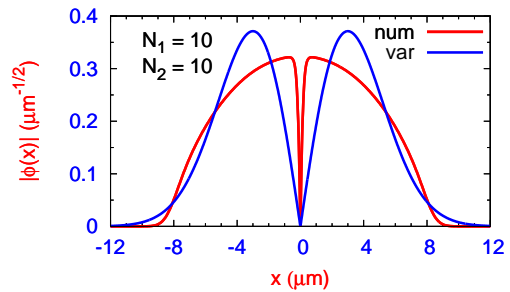}
\end{center}

\caption{(Color online)  The black 
soliton  $ |\phi(x)|$  
from the numerical solution of  Eq. (\ref{o})
(denoted num)  vs. $x$ 
for a DFFM with 
 $N_1=10$, $N_2=10$, $a_{12}/l=0.05 $, while   
$N_{11}=N_{22}\approx 275 $, and $N_{ij}=60$.
The corresponding variational solution obtained  from
Eq. (\ref{q}) (denoted var)
  is also shown. } 
\end{figure}

In our numerical study we take $l=1$ $\mu$m and 
consider a DFFM consisting of two electronic states of 
 $^{40}$K  atoms. For a BEC we consider  $^{39}$K  atoms.
This  corresponds to a radial
trap of frequency $\omega= \hbar/(l^2m) \approx 500$ Hz. 
Consequently, the  unit of
time is  $2/\omega \approx 4$ ms. 
For another fermionic atom the $\omega$ value 
gets changed accordingly for $l=1$ $\mu$m.

\subsection{Black Soliton in a BEC}

\begin{figure}
 
\begin{center}
\includegraphics[width=.9\linewidth]{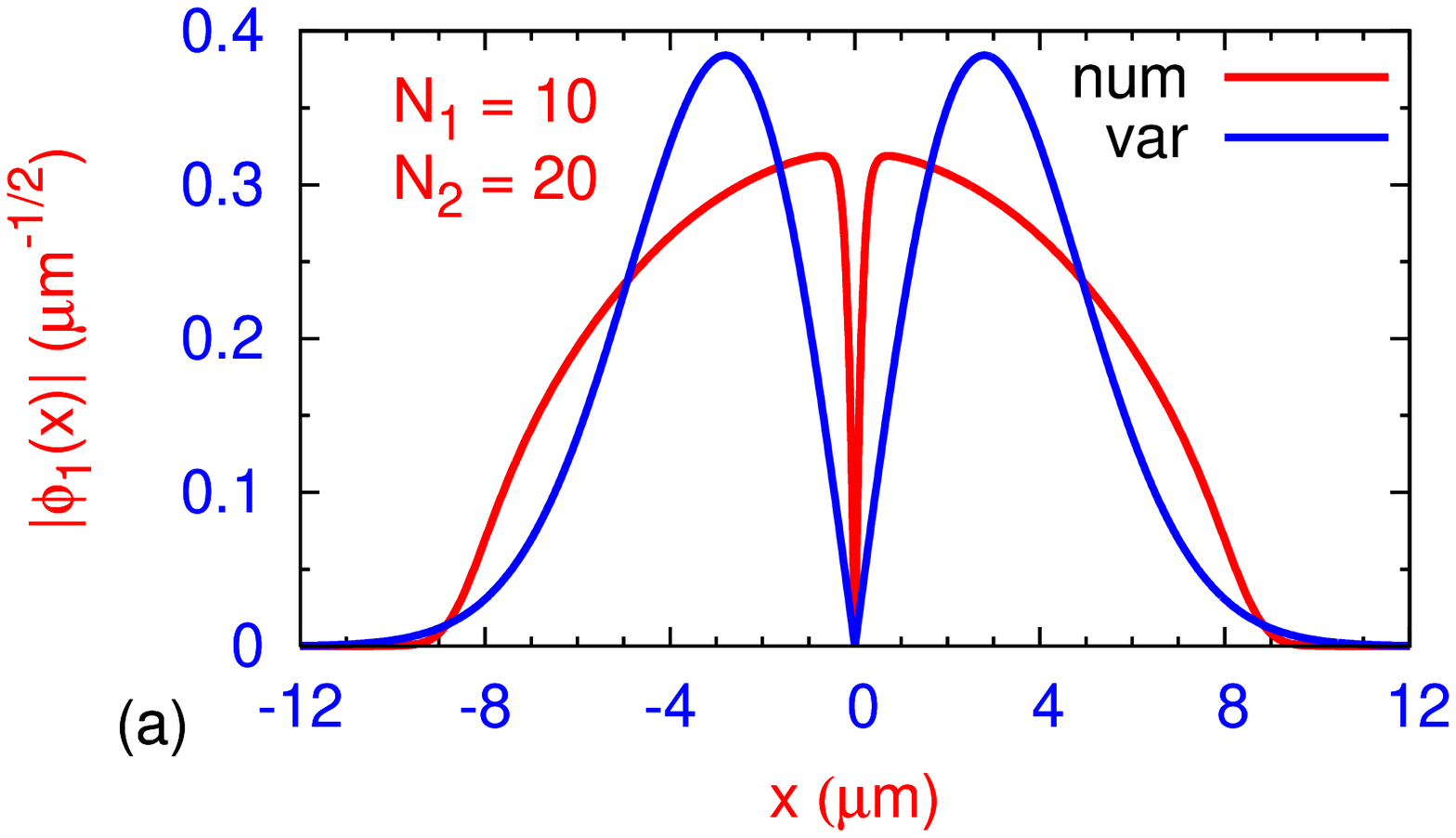}
\includegraphics[width=.9\linewidth]{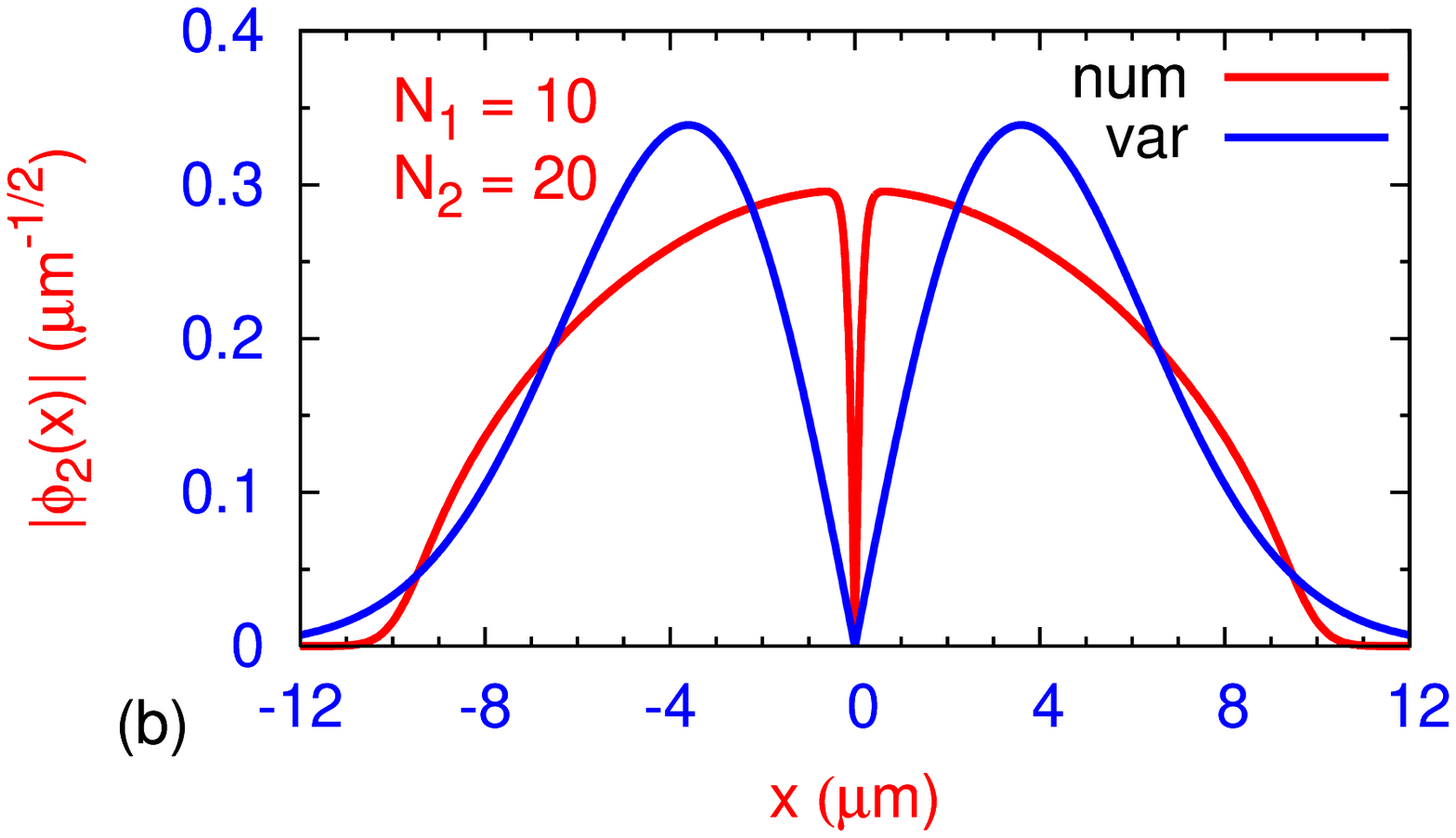}
\end{center}

\caption{(Color online)  The black solitons (a) $\phi_1(x)$ and
(b) $\phi_2(x)$ from the numerical solution of
 Eqs. (\ref{n}) (denoted num)
for a DFFM with $N_1=10$, $N_2=20$,
$a_{12}/l=0.05$, while ${\cal N}_{11}\approx 275$, ${\cal N}_{22}
\approx 436$,
${\cal N}_{12}=120$, and ${\cal N}_{21}=60$. The corresponding variational
solutions (denoted var) obtained from Eqs. (\ref{cp}) are also shown.}
\end{figure}

The results for black solitons in a BEC are obtained by solving
Eqs. (\ref{p}) and (\ref{q}) and are presented in Figs. 1 (a) and (b) for
$n=10$ and 100, respectively, where the soliton profiles are displayed. 
The variational profiles lie close to the numerical results and confirm
the existence of a black soliton in this case. For smaller
values of nonlinearity, the solitons have a probability distribution close
to the ansatz given by (\ref{v}). Consequently, 
the variational result is more accurate. For larger nonlinearities the
actual probability distribution obtained from the numerical
solution deviates from ansatz (\ref{v}) as can be seen
from Fig. 1 (b). By construction the variational ansatz (\ref{v}) is
stationary. The close agreement between the variational and numerical
solutions demonstrates the stationary nature of these dark solitons.  
Hence, contrary to comments in the literature \cite{ds2}, these black
solitons of
a trapped
BEC are  stationary.  
Although,
we illustrate the present simulation for a black soliton for a repulsive
BEC, stationary dark soliton also exists for an attractive BEC \cite{dsa}.

\subsection{Black Soliton in a DFFM}

The numerical results for black solitons in a DFFM are obtained by solving
Eqs. (\ref{o}) and (\ref{n}) for the symmetric ($N_1=N_2$) and
asymmetric
($N_1\ne N_2$) cases, respectively. The corresponding variational results
are obtained from Eqs. (\ref{sim}) and (\ref{cp}), respectively.
First we present the numerical and
variational 
results for the
symmetric situation with  $a_{12}/l=0.05$,  $\nu =0.1$, and 
$N_1=N_2=10$ in Fig. 2.   The nonlinearities here are quite large, e.g., 
$N_{jj}\approx 275$ and $N_{jk}= 60$, $j\ne k = 1,2$. The variational
and
numerical results in this case compare favorably as can be seen in Fig. 2.

Next we present the numerical and 
variational results for the asymmetric case 
with $N_1=10$ and $N_2=20$. $a_{12}/l=0.05$, and 
$\nu = 0.1$ corresponding to even larger nonlinearities 
${\cal N}_{11}\approx 275$, ${\cal N}_{22}\approx 436$,
${\cal N}_{12}= 120$,
and ${\cal N}_{21}= 60.$ The variational and numerical 
results in this case for black solitons for components 1 and 2 are
exhibited in Fig. 3 (a) and (b).

From Figs. 1$-$3 we see that  although the variational result is a good
quantitative approximation to the exact result for small values of
nonlinearity (Figs. 1), it ceases to be so for larger values of
nonlinearity (Figs. 2 and 3). This is true, as for large values 
of nonlinearity the harmonic-oscillator ansatz for the soliton
profile (\ref{v}) is not a good approximation to the accurate numerical
result.   In other words, the variational calculation confirms a
stationary black soliton in all cases. A poor initial variational
harmonic-oscillator ansatz for the soliton is responsible for 
the imprecise variational results for large nonlinearities.

\begin{figure}
 
\begin{center}
\includegraphics[width=.9\linewidth]{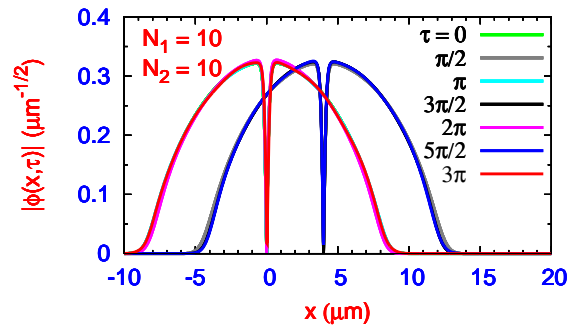}
\end{center}

\caption{(Color online)  The dark soliton profile for the 
black soliton of a DFFM of 
Fig. 2 with $N_1=N_2=10$ at times $\tau=$ $0, \pi/2, \pi, 3\pi/2, 2\pi,
5\pi/2, $ and $3\pi$ after the soliton has been set into oscillation by
displacing the trap through a distance of 2 $\mu$m.}

\end{figure}

\begin{figure}
 
\begin{center}
\includegraphics[width=.9\linewidth]{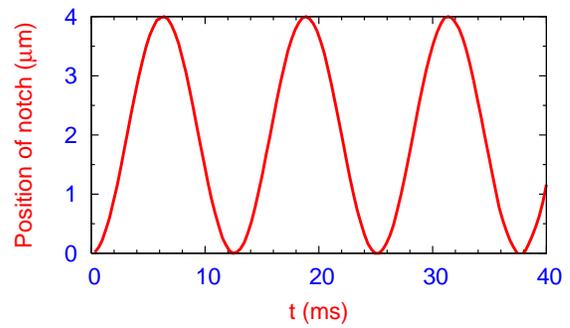}
\end{center}

\caption{(Color online)  The simple harmonic motion of the minimum of the
notch of the black soliton(s) 
of Fig. 2  after the soliton(s)  have 
been set into oscillation by displacing the trap through a distance of 2
$\mu$m. }
\end{figure}

To demonstrate the robustness of the black solitons, we introduced
a displacement in the notch of the black soliton upon formation 
by displacing the center of the harmonic trap by a small distance. 
The dark soliton(s) execute harmonic oscillation for small displacements 
during a moderate time interval. For the soliton of Fig. 2 with
$N_1=N_2=10$ the oscillation is illustrated  in Fig. 4 by exhibiting 
successive snapshots of the dark soliton at positions of maximum
displacement at times $\tau = 0, \pi/2, \pi, 3\pi/2, 2\pi, 5\pi/2$ and 
$3\pi$ 
for a trap displacement of 2 $\mu$m at time 0. The soliton remains black
at all times with the minimum at the notch equaling zero. The simple
harmonic motion of the position of the notch is illustrated in Fig. 5. The
angular frequency of oscillation as obtained from Fig. 5 is  the trap
frequency 500 Hz.  The numerical simulations presented in Figs. 4 and 5
demonstrate the stationary nature  of the black solitons under small
oscillations.

\begin{figure}
 
\begin{center}
\includegraphics[width=.9\linewidth]{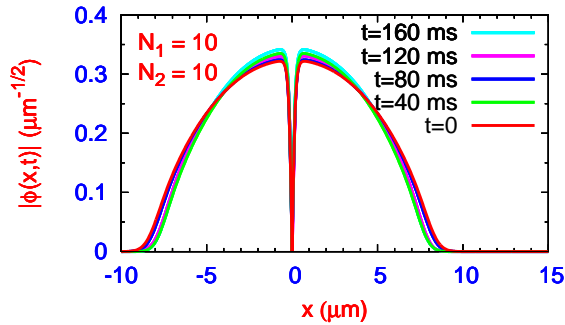}
\end{center}

\caption{(Color online)  The dark soliton profile for the black soliton of
Fig. 2 with $N_1=N_2=10$ at times $t=$ 0, 40 ms, 80 ms, 120 ms, and  160
ms after the soliton has been set into breathing oscillation by
increasing the strength of the trap by $10\%.$}

\end{figure}

To further demonstrate the stationary nature of the black solitons,
after their formation 
we increased suddenly the strength  of the harmonic trap by $10\%$: $x^2
\to 1.1 x^2$. The stability of the dark soliton for $N_1=N_2= 10$ under
this sudden change in the trap frequency is exhibited in Fig. 6 where we
plot the soliton profile at different times. The soliton executes
reasonably stable breathing oscillation. To study the breathing
oscillation in detail we calculate the root-mean-square (rms) size 
$\langle s \rangle$ 
of the
soliton at different times from a numerical solution of the
mean-field equation and the result is plotted in Fig. 
7 (a). We also
solved the variational equation (\ref{min}), an ordinary differential
equation by a fourth-order Runge-Kutta rule, to calculate the size of the
soliton and the time variation of the results is plotted in Fig. 7
(b). The amplitude and frequency of the variational oscillation are  in
good qualitative agreement with the precise numerical results. However,
there is quantitative disagreement between the two. After multiplying the  
variational size by 0.89, the results in Fig. 7 (b) are in good agreement 
with  the numerical result in Fig. 7 (a). The period  of breathing
oscillation from Fig. 7 (a) is 7.3 ms and that from Fig. 7 (b) is 
7.6 ms. The fair agreement between the variational and numerical results
of
breathing oscillation  illustrates that the variational description of the
stationary dark soliton is on the right track and confirms
the stationary nature of the black
soliton. 

\begin{figure}
 
\begin{center}
\includegraphics[width=.9\linewidth]{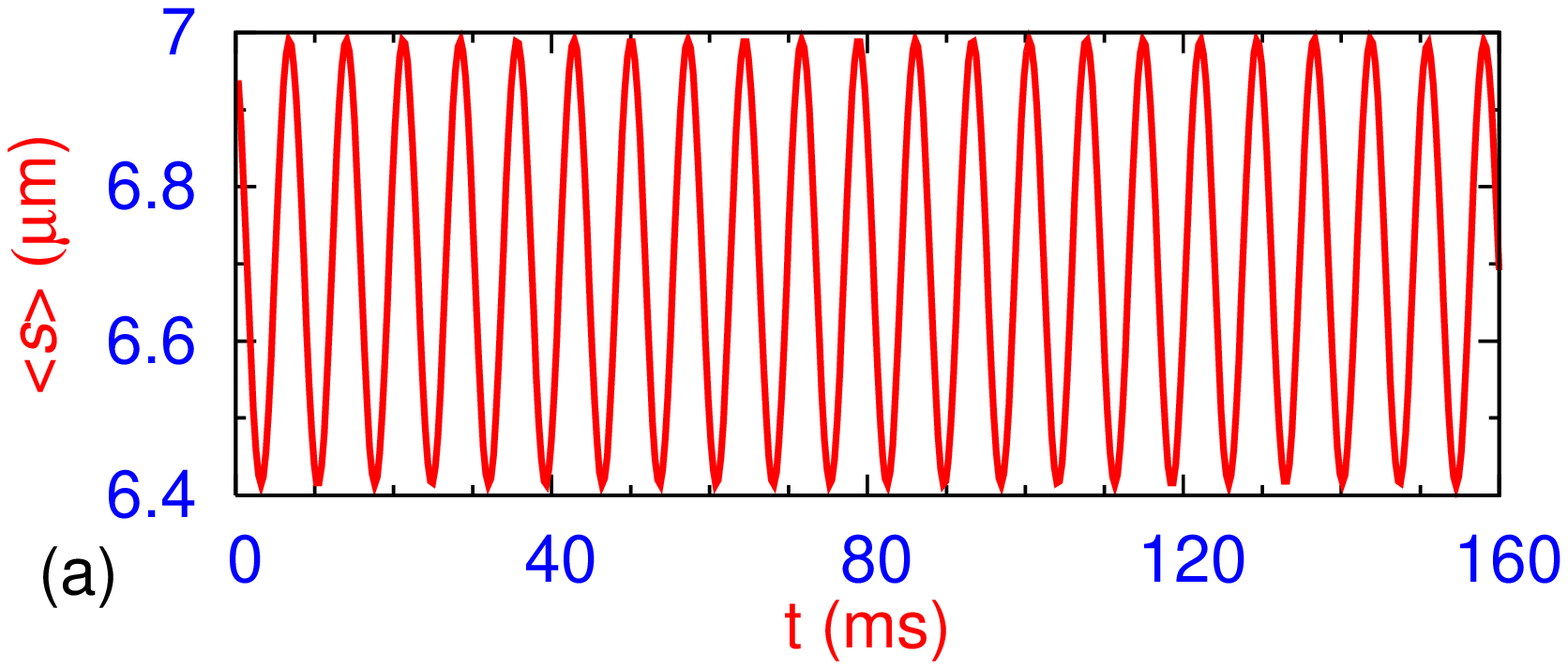}
\includegraphics[width=.9\linewidth]{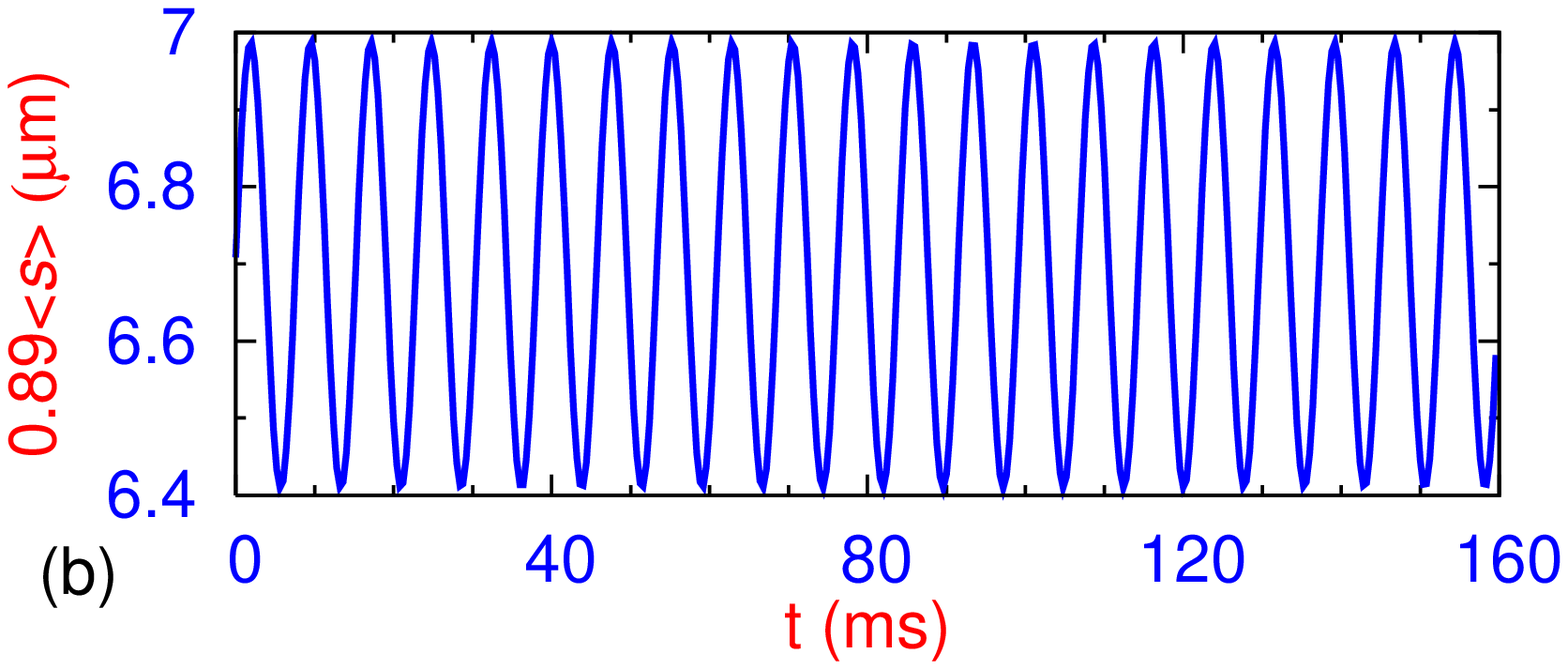}
\end{center}

\caption{(Color online)  
The periodic  motion of the rms size $\langle s\rangle$ of the dark
soliton after
suddenly 
increasing
the strength of the trap by 10 $\%$: (a) numerical result, (b) variational
result. }
\end{figure}

\section{Summary}

In conclusion, we used a coupled mean-field-hydrodynamic model for a DFFM
to study the formation of black  solitons in a
quasi-one-dimensional geometry by numerical and variational methods. The
existence of stationary dark solitons is predicted by variational study
and confirmed by detailed numerical analysis. 
The stationary nature 
of the present solitons is demonstrated numerically through
their (a) 
sustained oscillation initiated by a sudden displacement of the
harmonic trap, as well as (b) breathing oscillation initiated by a sudden
increase of the strength of the harmonic trap by $10\%$: $x^2 \to  1.1
x^2$.  Here we
used a set of mean-field equations for the DFFM.  A proper treatment of a
degenerate Fermi gas  or DFFM should be done using a fully antisymmetrized
many-body Slater
determinant wave function \cite{yyy} as in the case of
scattering
involving many electrons \cite{ps}. However, in view of the success of a
fermionic mean-field model 
in studies of collapse \cite{ska}, bright \cite{fbs2} 
and dark  \cite{fds} soliton 
in a DBFM and of mixing and demixing in a DFFM \cite{mix},
we
do
not believe that the present study on bright solitons in a DFFM to be so
peculiar as to have no general validity.

\acknowledgments

The work is 
supported in part by the CNPq and FAPESP
of Brazil.


\end{document}